\documentclass{IEEEtran}

\usepackage{subfigure}
\usepackage[cmex10]{amsmath}
\usepackage{amssymb}
\usepackage{epsfig}
\usepackage{cite}
\usepackage{graphicx}
\usepackage{color}
\usepackage{bm}
\usepackage{booktabs}
\usepackage{gensymb}
\usepackage{mathrsfs}
\usepackage{xfrac}
\usepackage{algorithm}
\usepackage{algorithmic}
\usepackage{multirow}
\usepackage{float}
\usepackage{units}
\newfloat{figtab}{htb}{fgtb}
\makeatletter
 \newcommand\figcaption{\def\@captype{figure}\caption}
  \newcommand\tabcaption{\def\@captype{table}\caption}

\newlength{\figwidth}


\begin{document}
\title{\LARGE Integrated Sensing and Communications for Pinching-Antenna\\ Systems (PASS)}
\author{Zheng Zhang, Zhaolin Wang, Xidong Mu  Bingtao He, Jian Chen, and Yuanwei Liu \vspace{-5mm}
\thanks{Zheng Zhang, Bingtao He, and Jian Chen are with the School of Telecommunications Engineering, Xidian University, Xi'an 710071, China (e-mail: zzhang\_688@stu.xidian.edu.cn; bthe@xidian.edu.cn; jianchen@mail.xidian.edu.cn).}
\thanks{Zhaolin Wang is with the School of Electronic Engineering and Computer Science, Queen Mary University of London, London E1 4NS, U.K. (e-mail: zhaolin.wang@qmul.ac.uk).}
\thanks{Xidong Mu is with Queen's University Belfast, Belfast, BT3 9DT, U.K. (email: x.mu@qub.ac.uk)}
\thanks{Yuanwei Liu is with the Department of Electrical and Electronic Engineering, The University of Hong Kong, Hong Kong (e-mail: yuanwei@hku.hk).}

}

\maketitle

\begin{abstract}
    An integrated sensing and communication (ISAC) design for pinching antenna systems (PASS) is proposed, where the pinching antennas are deployed to establish reliable line-of-sight communication and sensing links. More particularly, a separated ISAC design is proposed for the two-waveguide PASS, where one waveguide is used to emit the information-bearing signals for ISAC transmission while the other waveguide is used to receive the reflected echo signals. Based on this framework, a penalty-based alternating optimization algorithm is proposed to maximize the illumination power as well as ensure the communication quality-of-service requirement. Numerical results demonstrate that the proposed PASS-ISAC scheme outperforms the conventional antenna scheme.
\end{abstract}
\IEEEpeerreviewmaketitle

\section{Introduction}
Fuelled by the burgeoning demands for massive data transmission and pervasive network coverage, flexible antennas have emerged as a promising technique for sixth-generation (6G) cellular systems. Benefiting from their ability to reconfigure the wireless channel, flexible antennas can significantly enhance the throughput of wireless networks. However, traditional flexible antennas (e.g., movable antennas \cite{movable} and fluid antennas \cite{fluid}) merely permit the adjustment of the antenna position within a range of orders of magnitude comparable to the carrier wavelength. Against this backdrop, the pinching antenna has emerged \cite{pinching_docomo}, which is a type of dielectric waveguide-based leaky wave antenna. By applying dielectric particles to a particular point on the dielectric waveguide, a pinching antenna can be activated to establish EM radiation fields and form a communication area \cite{pinching_ding}. Then, the EM signal inside
the dielectric waveguide will be radiated from the pinching antenna to free space with a defined phase shift adjustment (referred to as the pinching beamformer). Notably, as the dielectric waveguide can be pinched at any position to radiate radio waves, the pinching antenna can flexibly move along the dielectric waveguide over a length of dozens of meters, thereby relocating to the closest position to the receiver and creating reliable LoS links.

To enable emerging applications, such as autonomous driving, extended reality, and the Metaverse, sensing functionality is recognized as an important indicator of future networks. In pursuit of this vision, the integrated sensing and communication (ISAC) technology has drawn significant attention recently \cite{ISAC}, which aims to leverage the cellular network hardware platforms and dedicated signal processing algorithms to achieve the incorporation of communication and sensing functionalities. Recently, it has been claimed that conducting ISAC transmission in the pinching antenna systems (PASS) can further upgrade the communication and sensing (C\&S) performance of the network \cite{pinching_yuanwei}. On the one hand, the pinching antenna can be flexibly repositioned to augment the echo signal energy. On the other hand, the wide-range mobility characteristic of pinching antennas results in an antenna aperture spanning dozens of meters. It inherently enables near-field sensing, e.g., the possibility of simultaneous angular and distance information estimation and even target velocity sensing, thereby offering a more comprehensive and accurate sensing of the surrounding environment. Nevertheless, as of the present moment, research in the PASS-ISAC remains conspicuously absent.

Motivated by the above, this paper proposes a separated ISAC design for PASS. To elaborate, the base station (BS) is connected with two dielectric waveguides, where one waveguide is used to transmit the downlink signals, while the other is employed to collect the reflected echo signals from the target. We aim to maximize the illumination power at the target while satisfying the quality-of-service (QoS) requirement of the communication user by optimizing the pinching beamforming offered by the mobility of pinching antennas. A penalty-based alternating optimization (AO) algorithm is proposed to handle the non-convex optimization problem, where the positions of pinching antennas are updated in an element-wise manner. Numerical results evaluate the superiority of the proposed scheme over the baseline schemes.

\section{System Model and Problem Formulation}\label{Section_2}

\begin{figure}[t]
  \centering
  \includegraphics[scale = 0.45]{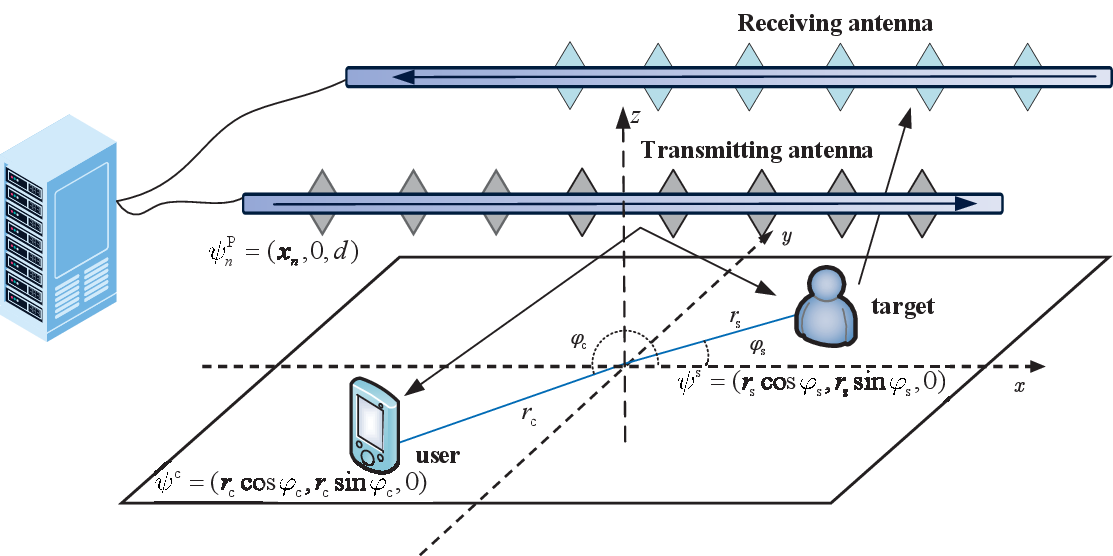}
  \caption{The separated ISAC design for PASS.}
  \label{model}
\end{figure}

As shown in Fig. \ref{model}, we consider a PASS-ISAC system, where a dual-function BS conveys with a single-antenna communication user while sensing a point-like target. The BS is connected with two dielectric waveguides of length $L$, each of which consists of $N$ pinching antennas. To achieve the simultaneous C\&S transmission, a separated ISAC design is proposed. Specifically, the downlink information-bearing signals are emitted from one waveguide (referred to as transmitting antennas). Then, the reflected echoes from the target would be collected at the other waveguide (referred to as receiving antennas), which are further transmitted to the BS for parameter estimation.

A three-dimensional (3D) coordination system is considered, where two dielectric waveguides extended from the BS are assumed to be parallel to the x-axis with respect to the x-o-y plane at a height $d$. The position of the $n$-th pinching antenna distributed along the transmitting and receiving dielectric waveguides can be denoted as $\psi_{n}^{\text{p}}=(x_{n}^{\text{p}},0,d)$ and $\psi_{n}^{\text{q}}=(x_{n}^{\text{q}},y^{\text{q}},d)$. The communication user and sensing target are located in the x-o-y plane. Let $r_{\text{c}}$ and $\varphi_{\text{c}}$ denote the distance and the azimuth angle of the communication user relative to the origin of the coordinate system. Thus, the coordinates of communication user is given by $\psi^{\text{c}}=(r_{\text{c}}\cos\varphi_{\text{c}},r_{\text{c}}\sin\varphi_{\text{c}},0)$. Similarly, the target is located in $\psi^{\text{s}}=(r_{\text{s}}\cos\varphi_{\text{s}},r_{\text{s}}\sin\varphi_{\text{s}},0)$. Furthermore, we assume the target is a static node or moves at a low speed. Thus, the Doppler effect is neglected in this work.

\subsection{Channel Model}
In the considered network, the pinching antennas are non-uniformly disposed on the dielectric waveguide covering the entire range of the user's activity, which implies that the aperture of the pinching antennas may have the same order of magnitude as the signal transmission distance. Without loss of accuracy, we adopt the spherical-wave-based near-field channel model, where only the LoS path is considered. Consequently, the distance from the $n$-th pinching antenna to the target is given by
\begin{align}
\nonumber
r_{n}^{\zeta}(r_{\zeta},\varphi_{\zeta}) &= \|\psi^{\zeta}-\psi_{n}^{\text{p}}\|\\ \label{NF_distance}
&= \sqrt{r_{\zeta}^{2}-2r_{\zeta}\cos\varphi_{\zeta}x_{n}^{\text{p}}
+(x_{n}^{\text{p}})^{2}+d^{2}},\ \zeta\in\{\text{s},\text{c}\},
\end{align}
Thus, the free space channel vector from the transmitting antennas to the target and the communication user can be expressed as
\begin{align}
\label{channel-model_sense}
\mathbf{h}_{\text{s}}(\mathbf{x}^{\text{p}})= \left[\frac{\eta^{\frac{1}{2}} e^{-\jmath \frac{2\pi}{\lambda}r_{1}^{\text{s}}(r_,\varphi_{\text{s}})}}
{r_{1}^{\text{s}}(r_{\text{s}},\varphi_{\text{s}})},\cdots,
\frac{\eta^{\frac{1}{2}} e^{-\jmath \frac{2\pi}{\lambda}r_{N}^{\text{s}}(r_{\text{s}},\varphi_{\text{s}})}}
{r_{N}^{\text{s}}(r_{\text{s}},\varphi_{\text{s}})}\right]^{H},
\end{align}
\begin{align}
\label{channel-model_com}
\mathbf{h}_{\text{c}}(\mathbf{x}^{\text{p}})= \left[\frac{\eta^{\frac{1}{2}} e^{-\jmath \frac{2\pi}{\lambda}r_{1}^{\text{c}}(r_,\varphi_{\text{c}})}}
{r_{1}^{\text{c}}(r_,\varphi_{\text{c}})},\cdots,
\frac{\eta^{\frac{1}{2}} e^{-\jmath \frac{2\pi}{\lambda}r_{N}^{\text{c}}(r_,\varphi_{\text{c}})}}
{r_{N}^{\text{c}}(r_,\varphi_{\text{c}})}\right]^{H},
\end{align}
where $\mathbf{x}^{\text{p}}=[x_{1}^{\text{p}},\cdots,x_{N}^{\text{p}}]$ denotes the coordinates of pinching antennas, $\lambda=\frac{c}{f_{\text{c}}}$ denotes the wavelength, $f_{\text{c}}$ is the frequency of the carrier wave, $\eta=\frac{c^{2}}{16\pi^{2}f_{\text{c}}^{2}}$, and $c$ denotes the speed of light.

In this paper, the BS aims to utilize the communication signal to achieve simultaneous communication and target sensing. Consider a coherent time block of length $T$, the communication channel condition and the sensing parameters are assumed to remain unchanged during one coherent time block. Thus, the emitted signal at the $t$-th time slot is given by  $s(t)\in\mathbb{C}$, which is assumed to be normalized and independently distributed, i.e., $\mathbb{E}\{|s(t)|^{2}\}=1$ and $\mathbb{E}\{s(t)s^{*}(\bar{t})\}=0$. On receiving $s(t)$, the dielectric waveguide radiates the signal $\mathbf{x}(t) = \sqrt{P_{\text{T}}}\mathbf{g}(\mathbf{x}^{\text{p}})s(t)$, where $\mathbf{g}(\mathbf{x}^{\text{p}})$ denotes the in-waveguide channel and can be expressed as
\begin{align}\label{emitted_signal}
\mathbf{g}(\mathbf{x}^{\text{p}}) = \left[\sqrt{\alpha_{1}}e^{-\jmath \theta_{1}},\cdots,\sqrt{\alpha_{N}}e^{-\jmath \theta_{N}}\right]^{T},
\end{align}
where $\theta_{n}$ denotes the radiation phase shift at the $n$-th pinching antenna, and $P_{\text{T}}$ denotes the transmit power at the BS. $\alpha_{n}$ denotes the power allocation coefficients at the $n$-th pinching antenna, which can be modeled as the equal power allocation model $\sqrt{\alpha_{n}}=\sqrt{\frac{\alpha_{\text{s}}}{N}}$ \cite{pinching_ding} or the proportional power allocation model $\sqrt{\alpha_{n}}=\delta(\sqrt{1-\delta^{2}})^{n-1}$ \cite{pinching_zhaolin}. $\delta=\sqrt{1-(1-\alpha_{\text{s}})^{\frac{1}{N}}}$ represents the proportional coefficient, and $\alpha_{\text{s}}=\sum_{n=1}^{N}\alpha_{n}$ denotes the radiation coefficient of pinching antennas. For ease of implementation, the equal power allocation model is considered in this paper.  $\theta_{n}$ is defined by $2\pi \eta_{\text{eff}}\frac{\|\psi_{0}^{\text{p}}-\psi_{n}^{\text{p}}\|}{\lambda}$, where $\psi_{0}^{\text{p}}$ denotes the location of the feed point, and $\eta_{\text{eff}}$ denotes the effective refractive index of the dielectric waveguide.

\subsection{Signal Model}
With the above channel model, it is readily observed that the positions of pinching antennas have a significant impact on both the free space channel $\{\mathbf{h}_{\text{s}}(\mathbf{x}^{\text{p}}),\mathbf{h}_{\text{c}}(\mathbf{x}^{\text{p}})\}$ and the in-waveguide channel $\mathbf{g}(\mathbf{x}^{\text{p}})$. As a result, it becomes possible to establish favorable wireless propagation while manipulating the radiated characteristics of signals by altering the positions of pinching antennas in the PASS. To characterize the two aspects of the signal reconfiguration capabilities of pinching antennas, we refer to it as \textbf{pinching beamforming} in this paper. Let $\mathbf{w}(\mathbf{x}^{\text{p}})$ and $\mathbf{v}(\mathbf{x}^{\text{p}})$ denote the pinching beamforming for the communication user and the sensing target, which are also the functions of $\mathbf{x}^{\text{p}}$. $\mathbf{w}(\mathbf{x}^{\text{p}})$ and $\mathbf{v}(\mathbf{x}^{\text{p}})$ are given by
\begin{align}
\label{com_pinching_beam}
\mathbf{w}(\mathbf{x}^{\text{p}})\!=\!\! \left[\!\frac{e^{-\jmath (\frac{2\pi}{\lambda}r_{1}^{\text{c}}(r_{\text{c}},\varphi_{\text{c}})\!+\! \theta_{1})}}{\frac{1}{\sqrt{\alpha_{1}}}r_{1}^{\text{c}}(r_{\text{c}},\varphi_{\text{c}})}\!,\!\cdots\!,\!\frac{
e^{-\jmath (\frac{2\pi}{\lambda}r_{N}^{\text{c}}(r_{\text{c}},\varphi_{\text{c}})\!+ \! \theta_{N})}}{\frac{1}{\sqrt{\alpha_{N}}}r_{N}^{\text{c}}(r_{\text{c}},\varphi_{\text{c}})}\!\right]^{T}\!\!,
\end{align}
\begin{align}
\label{sen_pinching_beam}
\mathbf{v}(\mathbf{x}^{\text{p}})\!=\!\! \left[\!\frac{ e^{-\jmath( \frac{2\pi}{\lambda}r_{1}^{\text{s}}(r_{\text{s}},\varphi_{\text{s}})\!+\!\theta_{1})}}
{\frac{1}{\sqrt{\alpha_{1}}}r_{1}^{\text{s}}(r_{\text{s}},\varphi_{\text{s}})}\!,\!\cdots\!,\!\frac{ e^{-\jmath( \frac{2\pi}{\lambda}r_{N}^{\text{s}}(r_{\text{s}},\varphi_{\text{s}})\!+\!\theta_{N})}}
{\frac{1}{\sqrt{\alpha_{N}}}r_{N}^{\text{s}}(r_{\text{s}},\varphi_{\text{s}})}\!\right]^{T}\!\!.
\end{align}
In this paper, we consider an ideal activation model of the pinching antenna, i.e., continuous activation. It indicates that the pinching antennas can be activated at any position of the dielectric waveguide. Thus, the positions of pinching antennas satisfy
\begin{align}
\label{continuous_activation}
\mathbf{x}^{\text{p}}\in\mathcal{X}=\left\{|x_{n}^{\text{p}}-x_{m}^{\text{p}}| \geq \Delta x\ (n\neq m),\ x_{n}^{\text{p}}\in \left[-\frac{L}{2},\frac{L}{2}\right]\right\},
\end{align}
where $\Delta x$ represents the minimum antenna space between two adjacent pinching antennas.

\textit{1) Communication Performance Metric:} With the aforementioned signal model, the received signals at the communication user are given by
\begin{align}
\nonumber
y(t) &=\sqrt{P_{\text{T}}}\mathbf{h}_{\text{c}}^{H}(\mathbf{x}^{\text{p}})
\mathbf{g}(\mathbf{x}^{\text{p}})s(t)+n(t) \\ \label{receive-signal-com}
&=\sqrt{P_{\text{T}}}\bm{\eta}^{H}\mathbf{w}(\mathbf{x}^{\text{p}})s(t)+n(t),
\end{align}
where $\bm{\eta}=[\eta^{\frac{1}{2}},\cdots,\eta^{\frac{1}{2}}]^{T}\in\mathbb{C}^{N \times 1}$ is a constant vector, and $n(t)\sim\mathcal{CN}(0,\sigma^{2})$ denotes the additive white Gaussian noise (AWGN) at the communication user. Hence, the achievable rate of the communication user is given by
\begin{align}
\label{rate-com}
R = \log_{2}\left(1+\frac{P_{\text{T}}|\bm{\eta}^{H}\mathbf{w}(\mathbf{x}^{\text{p}})|^{2}}{\sigma^{2}}\right).
\end{align}

\textit{2) Sensing Performance Metric:} For target sensing, we adopt the illumination power as the performance metric, which characterizes the received sensing signal power at the target \cite{Illumination_power1}.  Thus, the illumination power with respect to azimuth angle $\varphi_{\text{s}}$ and distance $r_{\text{s}}$ is given by
\begin{align}
\nonumber
P_{\text{s}} &= \mathbb{E}\left\{\left|\sqrt{P_{\text{T}}}\mathbf{h}_{\text{s}}^{H}(\mathbf{x}^{\text{p}})
\mathbf{g}(\mathbf{x}^{\text{p}})s(t)\right|^{2}\right\}\\ \label{beampattern}
&= P_{\text{T}}\bm{\eta}^{H}\mathbf{v}(\mathbf{x}^{\text{p}})\mathbf{v}^{H}(\mathbf{x}^{\text{p}})\bm{\eta}.
\end{align}

\subsection{Problem Formulation}
In this paper, we aim to maximize the illumination power $P(\theta_{\text{s}},r_{\text{s}})$ by designing the pinching beamformer, under the transmit power budget and communication QoS requirement, which is given by
\begin{subequations}
\begin{align}
\label{P1a} (\text{P}1)\quad  &\max\limits_{\mathbf{x}^{\text{p}}}    P_{\text{s}}\\
\label{P1b}\text{s.t.} \quad & R \geq R_{\text{QoS}},\\
\label{P1c}  & \mathbf{x}^{\text{p}}\in\mathcal{X},
\end{align}
\end{subequations}
where $R_{\text{QoS}}$ denotes the QoS requirement of the communication user. The problem (P1) is challenging to solve due to the quadratic objective function and the coupled variables.

\section{Pinching Beamforming Optimization}
In this section, we focus on the C\&S transmission design by optimizing the pinching beamforming. To deal with the coupled optimization variables, a penalty-based AO algorithm is proposed, where $\{\mathbf{x}^{\text{p}}\}$ is optimized in an element-wise manner.

To facilitate the optimization, we can rewrite the problem (P1) as
\begin{subequations}
\begin{align}
\label{P2a} (\text{P}2)\  \max\limits_{\mathbf{x}^{\text{p}}}\quad &   |\bm{\eta}^{H}\mathbf{v}(\mathbf{x}^{\text{p}})|^{2}\\
\label{P2b}\text{s.t.} \quad & |\bm{\eta}^{H}\mathbf{w}(\mathbf{x}^{\text{p}})|^{2} \geq \gamma_{\text{QoS}}\sigma^{2}, \\
\label{P2c} &  \eqref{P1c},
\end{align}
\end{subequations}
where $\gamma_{\text{QoS}}=\frac{2^{R_{\text{QoS}}}-1}{P_{\text{T}}}$.

In order to deal with the intractable objective and constraints, we consider a penalty-based two-layer framework. To elaborate, we introduce auxiliary variables $\mathbf{\tilde{w}}$ and $\mathbf{\tilde{v}}$ to replace $\mathbf{w}(\mathbf{x}^{\text{p}})$ and $\mathbf{v}(\mathbf{x}^{\text{p}})$, respectively. Thus, we have the equality constraints $\mathbf{\tilde{w}}=\mathbf{w}(\mathbf{x}^{\text{p}})$ and $\mathbf{\tilde{v}}=\mathbf{v}(\mathbf{x}^{\text{p}})$. By relocating the equality constraint to the objective function and serving as a penalty term, the problem (P2) can be equivalently rewritten as
\begin{subequations}
\begin{align}
\label{P3a} (\text{P}3)\  \max\limits_{\mathbf{x}^{\text{p}},\mathbf{\tilde{w}},\mathbf{\tilde{v}}}\quad &   |\bm{\eta}^{H}\mathbf{\tilde{v}}|^{2}-\frac{1}{2\varrho}\chi_{1}(\mathbf{x}^{\text{p}},\mathbf{\tilde{w}},\mathbf{\tilde{v}})\\
\label{P3b}\text{s.t.} \quad & |\bm{\eta}^{H}\mathbf{\tilde{w}}|^{2} \geq \gamma_{\text{QoS}}\sigma^{2}, \\
\label{P3c}  & |\mathbf{\tilde{w}}_{[n]}|^2 \leq \frac{1}{N r_{\text{min,c}}^2}, \\
\label{P3d}  &  |\mathbf{\tilde{v}}_{[n]}|^2 \leq \frac{1}{N r_{\text{min,s}}^2}, \\
\label{P3e} &  \eqref{P1c},
\end{align}
\end{subequations}
where $\chi_{1}(\mathbf{x}^{\text{p}},\mathbf{\tilde{w}},\mathbf{\tilde{v}})=\|\mathbf{\tilde{w}}-\mathbf{w}(\mathbf{x}^{\text{p}})\|+
\|\mathbf{\tilde{v}}-\mathbf{v}(\mathbf{x}^{\text{p}})\|$ and $\varrho$ denotes the scaling factor of the penalty terms. Note that to avoid the infinite objective value, we introduce constraints \eqref{P3c} and \eqref{P3d}, where $r_{\text{min,c}}=\sqrt{(r_{\text{c}}\sin \varphi_{\text{c}})^{2}+d^2}$ and $r_{\text{min,s}}=\sqrt{(r_{\text{s}}\sin \varphi_{\text{s}})^{2}+d^2}$ denote the lower bounds of the distances from an arbitrary pinching antenna to the communication user and target. The problem (P3) is equivalent to the problem (P1) as constraints \eqref{P3c} and \eqref{P3d} can be obtained from the \eqref{P1c}, which restricts pinching beamforming $\{\mathbf{\tilde{w}},\mathbf{\tilde{v}}\}$ to the feasible region.

To address the quadratic objective and constraints, we apply the SDR technique to rewrite the problem (P3) as follows.
\begin{subequations}
\begin{align}
\label{P4a} (\text{P}4)\quad &\max\limits_{\mathbf{x}^{\text{p}},\mathbf{\tilde{W}},\mathbf{\tilde{V}}}   \text{Tr}(\bm{\eta}\bm{\eta}^{H}\mathbf{\tilde{V}})-\frac{1}{2\varrho}\chi_{2}
(\mathbf{x}^{\text{p}},\mathbf{\tilde{W}},\mathbf{\tilde{V}})\\
\label{P4b}\text{s.t.} \quad &  \mathbf{\tilde{W}}_{[n,n]}\leq \frac{1}{N r_{\text{min,c}}^2}, \\
\label{P4c}  &  \mathbf{\tilde{V}}_{[n,n]}\leq \frac{1}{N r_{\text{min,s}}^2}, \\
\label{P4d} &\text{Tr}(\bm{\eta}\bm{\eta}^{H}\mathbf{\tilde{W}}) \geq \gamma_{\text{QoS}}\sigma^{2},\\
\label{P4e} &\text{rank}(\mathbf{\tilde{W}}) = 1,\ \text{rank}(\mathbf{\tilde{V}}) = 1,\\
\label{P4f} & \mathbf{\tilde{W}}\succeq \mathbf{0},\ \mathbf{\tilde{V}}\succeq \mathbf{0},\\
\label{P4g} &  \eqref{P1c},
\end{align}
\end{subequations}
where $\mathbf{W}(\mathbf{x}^{\text{p}})=\mathbf{w}(\mathbf{x}^{\text{p}})\mathbf{w}^{H}(\mathbf{x}^{\text{p}})$, $\mathbf{\tilde{W}}=\mathbf{\tilde{w}}\mathbf{\tilde{w}}^{H}$, $\mathbf{V}(\mathbf{x}^{\text{p}})=\mathbf{v}(\mathbf{x}^{\text{p}})\mathbf{v}^{H}(\mathbf{x}^{\text{p}})$, $\mathbf{\tilde{V}}=\mathbf{\tilde{v}}\mathbf{\tilde{v}}^{H}$, and $\chi_{2}(\mathbf{x}^{\text{p}},\mathbf{\tilde{W}},\mathbf{\tilde{V}})=\|\mathbf{\tilde{W}}-\mathbf{W}(\mathbf{x}^{\text{p}})\|_{F}+
\|\mathbf{\tilde{V}}-\mathbf{V}(\mathbf{x}^{\text{p}})\|_{F}$. To solve the problem (P4), we propose a penalty-based AO algorithm, which alternately optimizes $\{\mathbf{\tilde{W}},\mathbf{\tilde{V}}\}$ and $\{\mathbf{x}^{\text{p}}\}$ in the inner layer and updates $\varrho$ in the outer layer.

\subsubsection{Inner layer iteration---subproblem with respect to $\{\mathbf{\tilde{W}},\mathbf{\tilde{V}}\}$} With the fixed $\{\mathbf{x}^{\text{p}}\}$, the problem (P4) is reduced to
\begin{subequations}
\begin{align}
\label{P5a} (\text{P}5)\  \max\limits_{\mathbf{\tilde{W}},\mathbf{\tilde{V}}}\  &   \text{Tr}(\bm{\eta}\bm{\eta}^{H}\mathbf{\tilde{V}})-\frac{1}{2\varrho}\chi_{2}(\mathbf{x}^{\text{p}},\mathbf{\tilde{W}},\mathbf{\tilde{V}})\\
\label{P5b}\text{s.t.} \quad & \eqref{P4b}-\eqref{P4f}.
\end{align}
\end{subequations}
To handle the rank-one constraint, we introduce non-negative auxiliary variables $\{\varpi_{1},\varpi_{2}\}$ and employ the difference-of-convex (DC) relaxation method \cite{T.Jiang_DCP} to rewrite the \eqref{P4c} as
\begin{align}\label{DC}
\begin{cases}
\Re(\text{Tr}(\mathbf{\tilde{W}}^{H}(\mathbf{I}-\mathbf{\tilde{w}}_{\text{max}}
\mathbf{\tilde{w}}_{\text{max}}^{H})))\leq\varpi_{1},\\
\Re(\text{Tr}(\mathbf{\tilde{V}}^{H}(\mathbf{I}-\mathbf{\tilde{v}}_{\text{max}}
\mathbf{\tilde{v}}_{\text{max}}^{H})))\leq\varpi_{2},
\end{cases}\ i\in\{1,2\},
\end{align}
where $\mathbf{\tilde{w}}_{\text{max}}$ and $\mathbf{\tilde{v}}_{\text{max}}$ represent the eigenvectors corresponding to the maximum eigenvalues of $\mathbf{\tilde{W}}$ and $\mathbf{\tilde{V}}$, respectively. As a result, the problem (P5) can be transformed into
\begin{subequations}
\begin{align}
\label{P6a} (\text{P}6)\  \max\limits_{\mathbf{\tilde{W}},\mathbf{\tilde{V}},\varpi_{i}}\quad    &\text{Tr}(\bm{\eta}\bm{\eta}^{H}\mathbf{\tilde{V}})-\frac{1}{2\varrho}\chi_{2}
(\mathbf{x}^{\text{p}},\mathbf{\tilde{W}},\mathbf{\tilde{V}})-\sum_{i=1}^{2}\frac{1}{2\varrho_{i}}\varpi_{i}\\
\label{P6b}\text{s.t.} \quad & \varpi_{i} \geq 0,\ i \in \{1,2\},\\
\label{P6c} & \eqref{P4b}-\eqref{P4f},\eqref{DC},
\end{align}
\end{subequations}
where $\varrho_{i}$ denotes the scaling factor of $\varpi_{i}$. The problem (P6) is a convex problem and can be directly solved. Thus, the rank-one solution $\{\mathbf{\tilde{W}},\mathbf{\tilde{V}}\}$ can be obtained by carrying out the \textbf{Algorithm \ref{PTL}}.

\subsubsection{Inner layer iteration---subproblem with respect to $\{\mathbf{x}^{\text{p}}\}$}
Note that the equality constraint $\mathbf{\tilde{W}}=\mathbf{W}(\mathbf{x}^{\text{p}})$ and $\mathbf{\tilde{V}}=\mathbf{V}(\mathbf{x}^{\text{p}})$ are equivalent to $\mathbf{\tilde{w}}=\mathbf{w}(\mathbf{x}^{\text{p}})$ and $\mathbf{\tilde{v}}=\mathbf{v}(\mathbf{x}^{\text{p}})$. As a result, the problem (P6) can be transformed into
\begin{subequations}
\begin{align}\label{P7a}
(\text{P}7)\  \min\limits_{\mathbf{x}^{\text{p}}}\quad
& \|\mathbf{\tilde{w}}-\mathbf{w}(\mathbf{x}^{\text{p}})\|+\|\mathbf{\tilde{v}}-\mathbf{v}(\mathbf{x}^{\text{p}})\|\\
\label{P7b}\text{s.t.} \quad & \eqref{P1c}.
\end{align}
\end{subequations}
It is easy to notice that $x_{n}^{\text{p}}$ and $x_{m}^{\text{p}}$ ($n\neq m$) are separated in the objective function but coupled in the constraint \eqref{P1c}, which motivates us to adopt the element-wise optimization framework. Therefore, with the fixed $\{x_{1}^{\text{p}},\cdots,x_{n-1}^{\text{p}},x_{n+1}^{\text{p}},\cdots,x_{N}^{\text{p}}\}$, the subproblem with respect to $x_{n}^{\text{p}}$ is given by
\begin{subequations}
\begin{align}
\nonumber (\text{P}8)\  \min\limits_{x_{n}^{\text{p}}}\quad
& \left|\mathbf{\tilde{w}}_{[n]}-\frac{e^{-\jmath (\frac{2\pi}{\lambda}r_{n}^{\text{c}}(r_{\text{c}},\varphi_{\text{c}})+\theta_{n})}}
{\sqrt{N}r_{n}^{\text{c}}(r_{\text{c}},\varphi_{\text{c}})}\right|\\
\label{P8a}&+\left|\mathbf{\tilde{v}}_{[n]}-\frac{e^{-\jmath (\frac{2\pi}{\lambda}r_{n}^{\text{s}}(r_{\text{s}},\varphi_{\text{s}})+\theta_{n})}}
{\sqrt{N}r_{n}^{\text{s}}(r_{\text{s}},\varphi_{\text{s}})}\right|\\
\label{P8b}\text{s.t.} \quad &  x_{n-1}^{\text{p}}+\Delta x \leq x_{n}^{\text{p}}\leq x_{n+1}^{\text{p}}-\Delta x,\\
\label{P8c} & \frac{-L}{2} \leq x_{n}^{\text{p}}\leq \frac{L}{2},
\end{align}
\end{subequations}
Then, the optimal $x_{n}^{\text{p}}$ can be obtained by the low-complexity one-dimensional search.

\subsubsection{Outer layer iteration} In the outer layer, we initialise a large $\varrho$ and update $\varrho$ at each outer iteration by
 \begin{align}\label{outer_update}
\varrho = \varrho \bar{c}_{2},
\end{align}
where $0<\bar{c}_{2}<1$ is the iteration coefficient of the penalty terms. The penalty-based AO algorithm is summarized in \textbf{Algorithm \ref{PAO}}.

\begin{algorithm}[t]
    \caption{Iterative algorithm for rank-one solution.}
    \label{PTL}
    \begin{algorithmic}[1]
        \STATE{Initialize $\mathbf{\tilde{v}}_{\text{max}}$, and $\mathbf{\tilde{w}}_{\text{max}}$. Set a convergence accuracy $\epsilon_{1}$.}
        \REPEAT
        \STATE{ update $\{\mathbf{\tilde{W}},\mathbf{\tilde{V}},\varpi_{i}\}$ by solving the problem (P6).}
        \STATE{ update the eigenvectors $\{\mathbf{\tilde{w}}_{\text{max}},\mathbf{\tilde{v}}_{\text{max}}\}$.}
        \STATE{ update $\varrho_{i}=\varrho_{i}\bar{c}_{1}$ ($0<\bar{c}_{1}<1$).}
        \UNTIL{ $\sum_{i=1}^{2}\varpi_{i}$ falls below a threshold of $\epsilon_{1}$.}
    \end{algorithmic}
\end{algorithm}
\begin{algorithm}[t]
    \caption{Penalty-based AO algorithm.}
    \label{PAO}
    \begin{algorithmic}[1]
        \STATE{Parameter Initialization. Set the convergence accuracy $\epsilon_{2}$ and $\epsilon_{3}$.}
        \REPEAT
        \REPEAT
        \STATE{ update $\{\mathbf{\tilde{w}},\mathbf{\tilde{v}}\}$ by carrying out \textbf{Algorithm \ref{PTL}}.}
        \STATE{ update $\mathbf{x}_{\text{p}}$ via the element-wise optimization.}
        \UNTIL{the objective value converges with an accuracy of $\epsilon_{2}$.}
        \STATE{update $\varrho=\varrho\bar{c}_{2}$ ($0<\bar{c}_{2}<1$).}
        \UNTIL{$\|\mathbf{\tilde{W}}-\mathbf{W}(\mathbf{x}^{\text{p}})\|_{F}+
\|\mathbf{\tilde{V}}-\mathbf{V}(\mathbf{x}^{\text{p}})\|_{F}\leq\epsilon_{3}$.}
    \end{algorithmic}
\end{algorithm}

The proposed penalty-based AO algorithm is summarized in \textbf{Algorithm \ref{PAO}}, which is assured to converge at least to a stationary point solution. The computational complexity of \textbf{Algorithm \ref{PAO}} mainly depends on solving the SDP problems (P6) and the one-dimensional exhaustive search. It is given by $\mathcal{O}\Big(\log(\frac{1}{\epsilon_{3}})
\log(\frac{1}{\epsilon_{2}}) \big[ \log(\frac{1}{\epsilon_{1}})N^{3.5}
+N\bar{Q}\big]\Big)$ \cite{SDR}, where $\bar{Q}$ represents the number of the quantization bits during the one-dimensional exhaustive search.

\section{Numerical Results}

This section evaluates the performance of the proposed PASS-ISAC framework. A 3D topological network setup is considered, where the dielectric waveguide is located in the x-o-z plane with a height of $d$ and a length of $50$ m. The communicating user and the sensing target are located in a square region centered at the origin in the x-o-y plane. Unless otherwise specified, the default simulation parameters are set as: $\sigma^{2}=-105$ dBm, $f=28$ GHz, $d=10$ m, $r_{\text{s}} = 30$ m, $\varphi_{\text{s}} = \frac{\pi}{3}$, $r_{\text{c}} = 15\sqrt{2}$ m, $\varphi_{\text{c}} = \frac{5\pi}{4}$, $N = 16$, $\eta_{\text{eff}} = 1.4$, $R_{\text{QoS}} = 10$ bps/Hz, $\epsilon_{1}=\epsilon_{2}=\epsilon_{3}=\epsilon_{4}=10^{-3}$, and $\alpha_{\text{s}}=1$. The other network parameters are shown in the captions of the figures.

\begin{figure}[t]
  \centering
  \includegraphics[scale = 0.45]{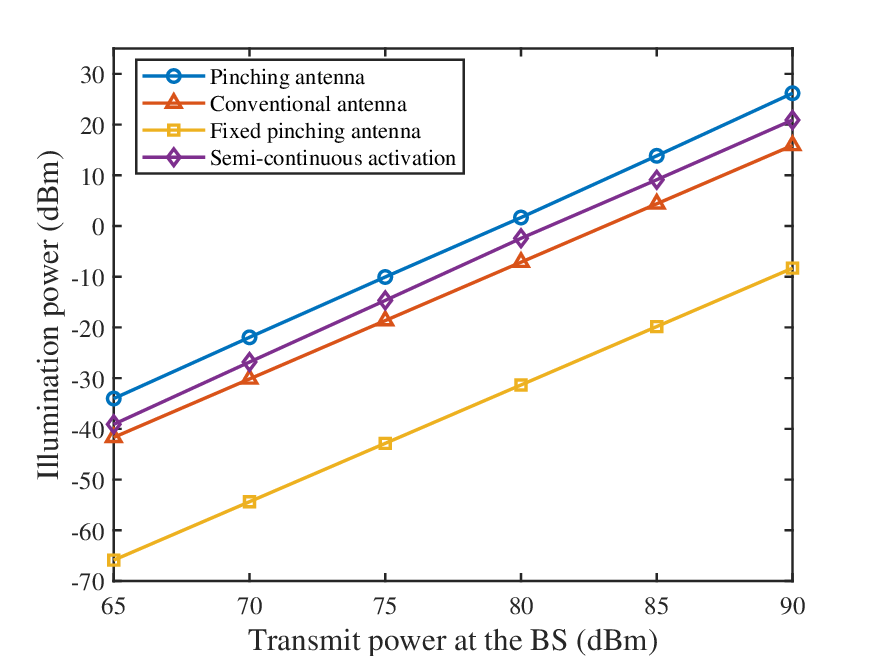}
  \caption{The illumination power versus the transmit power at the BS.}
  \label{Benchmark}
\end{figure}

To validate the performance of the proposed scheme, the following baseline schemes are considered in this paper:
\begin{itemize}
  \item \textbf{Conventional antenna}: In this scheme, we deploy $N$ conventional uniform linear array (ULA) at the BS as the transmitting antenna with an antenna spacing of $\frac{\lambda}{2}$. For fairness comparison, the transmitting antennas are connected to one RF chain and each antenna is associated with an analog phase shifter, which can be varied from $0$ to $2\pi$.
  \item \textbf{Fixed pinching antenna}: In this scheme, $N$ pinching antennas are uniformly spread along the dielectric waveguide, where the in-waveguide and free-space channels are determined by the fixed positions of the pinching antennas.
  \item \textbf{Semi-continuous activation}: In the semi-continuous activation scheme, we assume there are $N$ pinching antennas uniformly distributed along the dielectric waveguide, which are predetermined and cannot be changed. However, the pinching antennas are allowed to be adjusted in a small-scale range to alter the phase-shift response of the pinching beamforming, which has a negligible impact on the large-scale path loss.
\end{itemize}

In Fig. \ref{Benchmark}, we can observe that the pinching antenna achieves the highest illumination power compared to the other baseline schemes. This result can be expected because, compared with the baseline schemes, pinching antennas can be flexibly repositioned to attenuate the large-scale path loss between the pinching antennas and the receiving ends. Thus, more spatial degrees-of-freedom (DoFs) are provided to favor the communication and sensing performance. On the other hand, although the semi-continuous activation scheme cannot reduce the path loss by adjusting the antenna position over a wide range, it exhibits superior performance to the conventional antenna scheme because pinching antennas are spread over the entire communication/sensing area, which averagely closer to the receiving ends.

\begin{figure}[t]
  \centering
  \includegraphics[scale = 0.45]{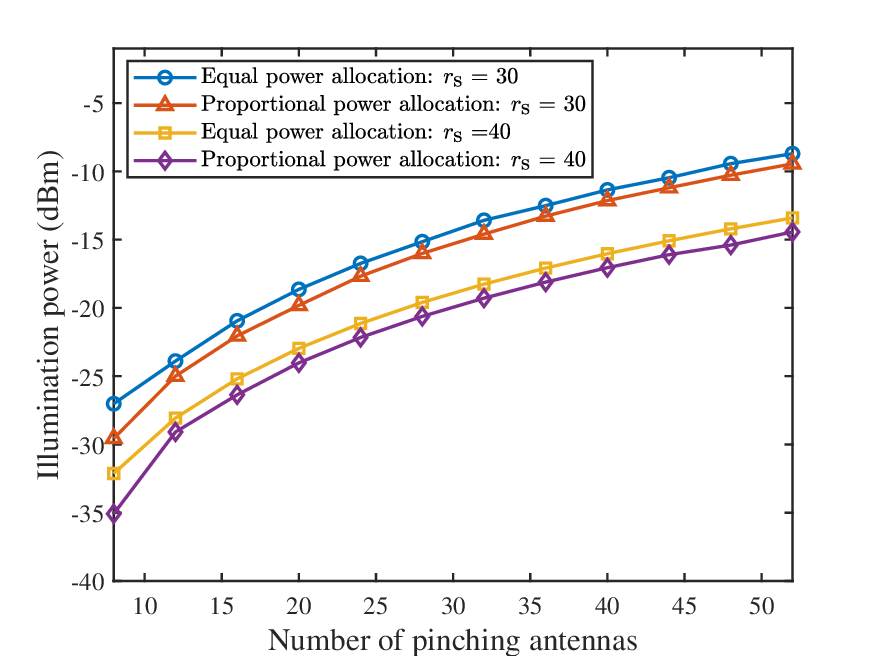}
  \caption{The illumination power versus the rotation angle of the dielectric waveguide, where $P_{\text{T}} = 70$ dBm.}
  \label{antenna_number}\vspace{-3mm}
\end{figure}

Fig. \ref{antenna_number} depicts the relationship between the illumination power and the number of activated pinching antennas, with a comparison of the proportional power allocation model. For fairness comparison, $\alpha_{\text{s}}=0.9$ for two power allocation models. As can be observed, the illumination power increases as the number of pinching antennas increases, which is because an increasing number of pinching antennas can improve the beam resolution and reduce the power leakage in irrelevant regions, thereby raising the illumination power at the target. It is also observed that the proportional power allocation is slightly inferior to the equal power allocation model, which verifies the effectiveness of the pinching antennas based on proportional power allocation model in reconfiguring signal propagation.

\begin{figure}[t]
  \centering
  \includegraphics[scale = 0.45]{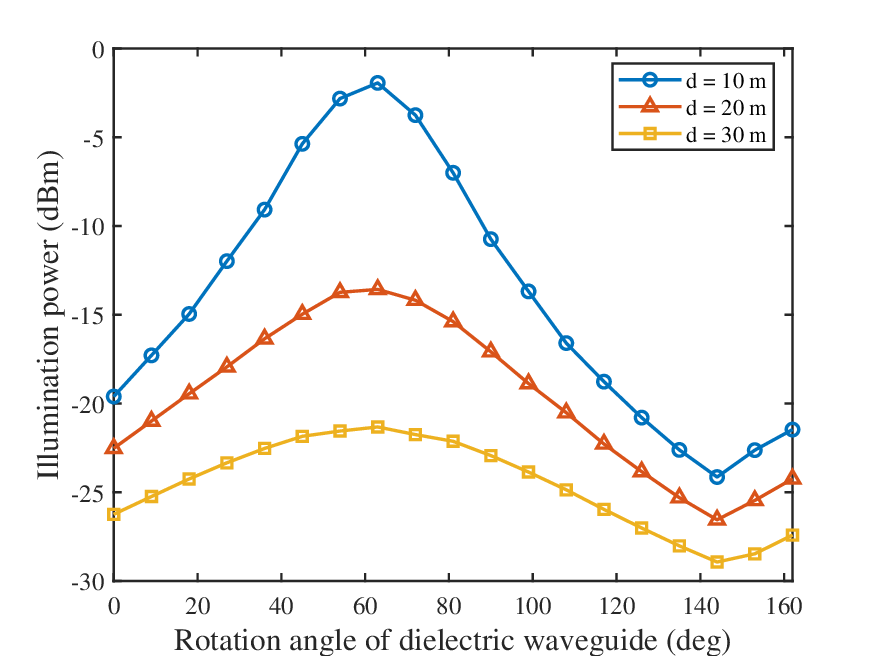}
  \caption{The illumination power versus the rotation angle of the dielectric waveguide, where $P_{\text{T}} = 70$ dBm.}
  \label{versus_angle}\vspace{-3mm}
\end{figure}
Fig. \ref{versus_angle} investigates the impact of the rotation angle of the dielectric waveguide on illumination power at the target. Here, we assume the dielectric waveguide can be rotated in a clockwise direction parallel to the x-o-y plane, where the rotation angle is defined as the angle entwined by the dielectric waveguide and the x-axis. From Fig. \ref{versus_angle}, it is shown that the illumination power first increases and then decreases as the rotation angle grows. This is due to the fact that when the rotation angle is $60^{\circ}$, the target is located underneath the dielectric waveguide, and it receives the maximal illumination power. As the rotation angle further rises, the distance between the target and the pinching antenna becomes large, so the illumination power gradually decreases. In addition, raising the height of the dielectric waveguide increases the average distance from the pinching antennas to the user and target, thus, the illumination power decreases as $d$ increases.

\section{Conclusion}
A novel PASS-ISAC framework has been proposed, where the pinching beamforming was exploited to realize the simultaneous C\&S transmission. A separated ISAC design was proposed for the two-waveguide PASS. A penalty-based AO algorithm was proposed to maximize the illumination power at the target while guaranteeing the QoS requirement of the communication user. Simulation results were provided to verify the superiority of the proposed PASS-ISAC framework over the other baseline schemes.

\vspace{-3mm}
\bibliographystyle{IEEEtran}
\bibliography{bib}
\end{document}